\documentclass[%
 reprint,
 amsmath,amssymb,
 aps,
]{revtex4-2}
\usepackage{graphicx}
\usepackage{dcolumn}
\usepackage{bm}
\usepackage{hyperref}

\DeclareUnicodeCharacter{2212}{-}

\begin{document}

\preprint{APS/123-QED}

\title{Cavity based non-destructive detection of photoassociation in a dark MOT }

\author{V. I. Gokul}
\author{Arun Bahuleyan}%
\author{S. P. Dinesh}
\author{V. R. Thakar}
\author{S. A. Rangwala}
\affiliation{%
 Raman Research Institute, C.V. Raman Avenue, Sadashivanagar, Bangalore 560080, India 
}%

\date{\today}

\begin{abstract}
The photoassociation (PA) of rubidium dimer (Rb${}_2$) in a dark magneto-optic trap (MOT) is studied using atom-cavity collective strong coupling. This allows non-destructive detection of the molecule formation process as well as rapid and repeated interrogation of the atom-molecule system. The vacuum Rabi splitting (VRS) measurements from the bright MOT are carefully calibrated against equivalent measurements with fluorescence. Further loading rates in dark MOT are determined using VRS. This method provides a reliable, fast, and non-destructive detection scheme for ultracold molecules when the atoms are non-fluorescing using the free atoms coupled to a cavity.   
\end{abstract}

\maketitle

\section{\label{sec:level1} Introduction}

\begin{figure}[t!]
    \centering
    \includegraphics[width=0.47\textwidth]{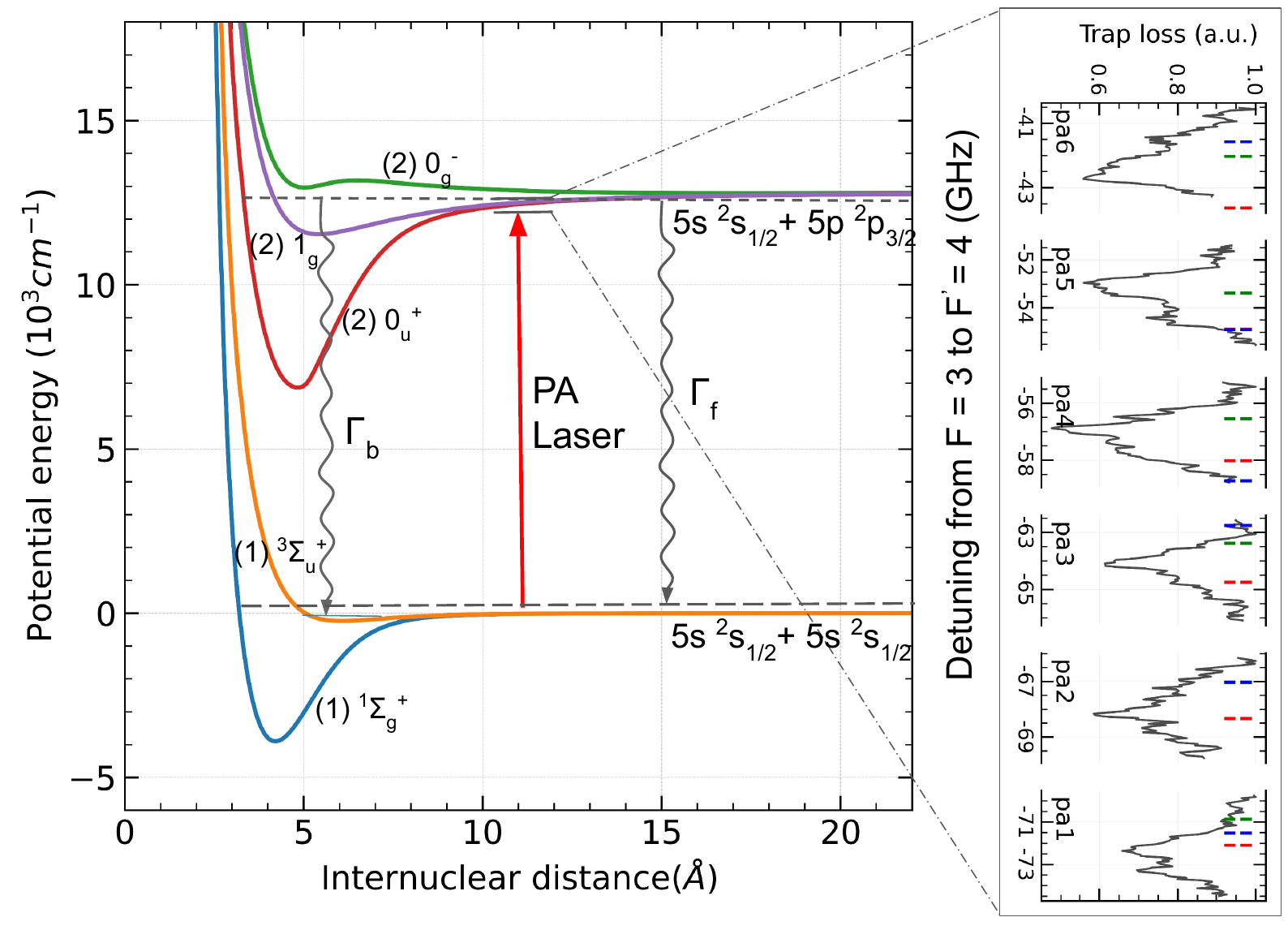}
    \caption{Potential energy curves for Rb${}_2$ molecule relevant in the experiment and various transitions. b) Trap-loss spectrum for different PA transitions in bright MOT. Here, PA frequency is referenced to $F=3$ to $F'=4$ D$2$ transition of Rb. }
    \label{fig:PEC}
\end{figure}

The detection of multiparticle interaction is a challenging experimental problem. In an ensemble, this can be done by detecting the post-collision product or by interrogating individual participants. This methods only allow partial information about the process of interest. For atomic systems, cavity-based measurements are very precise for studying atom-field interactions. In this work, we adapt cavity QED technique to detect the photoassociation (PA) of atoms. We aim to demonstrate the power of cavity based technique with PA as a representative example, though this technique has wider application, for instance, collision rate measurement between the two overlapping ensembles in the cavity mode volume. PA results when three particles, two free atoms in a scattering state absorb one photon and form a bound, excited molecule~\cite{Thorsheim,Jones}(Fig.~\ref{fig:PEC}). The absorbed photon is resonant with the free-to-bound transition, and the excited state molecule can subsequently decay into free atoms by emitting a photon $(\Gamma_f)$ or into a bound, ground-state molecule in one of many possible ro-vibrational states $(\Gamma_b)$. PA is efficient in an ensemble of ultracold atoms, where the energy of the scattering state is well-defined, favoring resonant excitation by a single, narrow-band laser. Suitable conditions for PA, therefore, require trapped atoms in magneto-optical traps (MOT's)~\cite{Dalibard89,Takekoshi,Lett93,Wang96,Fioretti1999,Bergeman_2006}, magnetic traps~\cite{Wang2004,Prodan}, or dipole traps~\cite{Miller1993,Staanum,Zahzam}.

Detection of PA in a bright MOT (where the MOT is observable in fluorescence) is done by measuring the resonant loss of the fluorescing atoms. Fluorescence detection is not an option in a dark MOT~\cite{Ketterle} and for other dark traps. Here, resonant multi-photon ionization (ReMPI) of the resulting ground state molecule and the subsequent detection of the molecular ion is the main detection tool available~\cite{Huang_2006, Comparat2000, Wang2004, Sage, Fioretti2001}. ReMPI has limited efficiency principally because the ground-state molecules produced by PA are created in various states due to spontaneous emission from the excited state.

 In this paper, we present cavity-based, non-destructive detection of the atoms, which are collectively strongly coupled to a low finesse cavity, during PA. In the experiment, we probe the vacuum Rabi splitting (VRS) of the atoms coupled to the cavity~\cite{Raizen,thompson1992}. As will be shown below, the technique works very well for a dark MOT, where atomic fluorescence is not detectable. In addition, this measurement is non-destructive, as it probes the atoms and not the molecules, with very low light intensity detuned with respect to the atomic resonance, resulting in no additional loss of atom from the trap. This combination of properties allows for continuous monitoring of PA, which we demonstrate so that its temporal evolution can also be measured and studied. Below, we first explain our experimental setup, calibrate our technique with a bright MOT, demonstrate results with the dark MOT, discuss the different ratios of bright and dark MOT resonance intensities, and conclude with some applications of this method. Our technique complements the recent direct strong coupling signature of PA~\cite{Konishi2021} observed in a high finesse $(F \sim 5\times10^4)$ cavity. Even though the finesse in our experiment doesn't allow for a direct strong coupling signature of PA, detecting atoms instead of molecules allows us to detect PA and calculate rates. A relatively low finesse $(\sim 330)$ cavity used in the experiment implies that one can perform these measurements even by having a cavity outside the vacuum.

\section{\label{setup} Experimental setup}

The experimental setup consists of a Fabry-Perot (FP) cavity co-centered with a dilute gas of ultracold ${}^{85}$Rb atoms that are trapped in an MOT. A detailed overview of the experimental setup is discussed in previous work~\cite{Tridib,Niranjan}. The FP cavity has length $L \approxeq 45.7 $ mm, finesse $F \sim 330$, and linewidth $\kappa \sim 4.4 $ MHz. A cooling beam detuned $-12$ MHz from $F=3$ to $F^{\prime}=4$ and repumper on $F=2$ to $F^{\prime}=3$ forms the bright MOT. The dark MOT is implemented by obstructing the center of an independent repumper beam with $\approx 2$ mm disk shelving the atoms in the $F=2$ non-fluorescing state. A double pass acousto-optic modulator (AOM) is used to switch both repumping lasers, allowing us to change between bright and dark MOT in our system. Additionally, a weak depumping beam $(\sim 1 \mu$W$)$ from $F=3$ to $F^{\prime}=3$ is used to maximize dark MOT density. For a cooling power of $25$ mW and repumping power of $3.5$ mW measured peak density of dark MOT, $\rho_0 \approx 1.7 \times 10^{10}$ cm${}^{-3}$. For all dark MOT measurements, the cavity is referenced to $F=2$ to $F^{\prime}=3$ transition~\cite{dinesh}, and a weak probe beam is scanned across the same transition. The cavity output is monitored using a photomultiplier tube (PMT) and a charge-coupled device (CCD) camera. A separate free-space setup is used to collect MOT fluorescence~\cite{Niranjan}. The PA beam has $\sim 2$ mm diameter and is incident perpendicular to the cavity axis. The wavelength of this PA laser is continuously monitored using a wavemeter.

In the experiment, $\kappa$ and excited state decay $(\Gamma)$ are much greater than single atom-cavity coupling constant $g_0 = \sqrt{\mu_{23}^2\omega_{23}/{2\hbar\epsilon_{0}V_c}} \approx 0.141$ MHz for $F=2 \rightarrow F'=3$ transition~\cite{Sanchez}, where $\mu_{23}$ is the transition dipole matrix element for probe transition and $V_c$ is the cavity mode volume. So, while a single atom cannot couple strongly to the cavity~\cite{Childs,Raizen} when $N_c$ atoms are present in $V_c$, each of these atoms can couple to a single cavity photon, resulting in an effective coupling strength of $g_N =  g_0 \sqrt{N_c}$~\cite{Agarwal,zhu,Raizen,Dutta}. In this collective strong coupling regime, the condition for observing strong coupling effects becomes $g_N = g_0\sqrt{N_c} \gg \Gamma,\kappa$. For a MOT with density distribution $ \rho(x,y,z) $ co-centered with a cavity mode, $N_c$ can be calculated using $N_c = \int \rho(x,y,z) |\psi(x,y,z)|^2 dV$, where $\psi$ is the cavity mode function~\cite{Niranjan}. 

For our experimental parameters, atom numbers as low as $N_c \gtrsim 1000$ can be measured via VRS in a collective strong coupling regime. Since the separation between VRS peaks is given by $2g_0\sqrt{N_c}$, any change in VRS corresponds to a change in $N_c$, which is a direct measure of change in MOT atom number~\cite{Dutta}. This allows detection of PA transitions in dark MOT using VRS. 

\section{\label{result} Results}

\subsection{\label{pa_bmot} PA resonances and loading rates in bright MOT}

The potential energy curves (PEC) for long-range photoassociated Rb${}_2$ ~\cite{Allouche,jyothi} molecule in D2 transition are shown in fig.~\ref{fig:PEC}. From the selection rule only transitions relevant and observed in our experiments are $0_g^-$, $0_g^+$ and $1_g$ ~\cite{Cline}. In the presence of a PA laser, two ground state Rb atoms can form a weakly bound excited state Rb${}_2$ molecule ($0_g^-$, $0_g^+$ or $1_g$). This excited state molecule decays back into a high vibrational ground state molecule (${}^1\Sigma_g^{+}$ or ${}^3\Sigma_u^{+}$) or two free atoms by photoemission. 

In the experiment, bright MOT has a peak density of $\sim 8 \times 10^{10}$ cm${}^{-3}$ and FWHM of $\sim 180$ $\mu$m. The PA laser is tuned from $12814$ cm${}^{-1}$ to $12816$ cm${}^{-1}$ and six prominent PA resonances are addressed. These transitions are labeled as pa1-pa6. The peak intensity of the PA laser is $\sim 15$ W$/$cm${}^2$, which is sufficient to saturate the PA transition. Fig.~\ref{fig:PEC} (b) shows the trap-loss measurement for these PA transitions. The dashed lines in the figure represent the closest calculated vibrational level for the excited state ab initio molecular potentials~\citep{jyothi}. As this method is non-destructive, one can perform continuous measurements to study the evolution and dynamics of the system. As proof of principle, we measure the loading rates in MOT using VRS and calculate loss rates in the system during PA.

\begin{figure}[t!]
    \centering
    \includegraphics[width=0.47\textwidth]{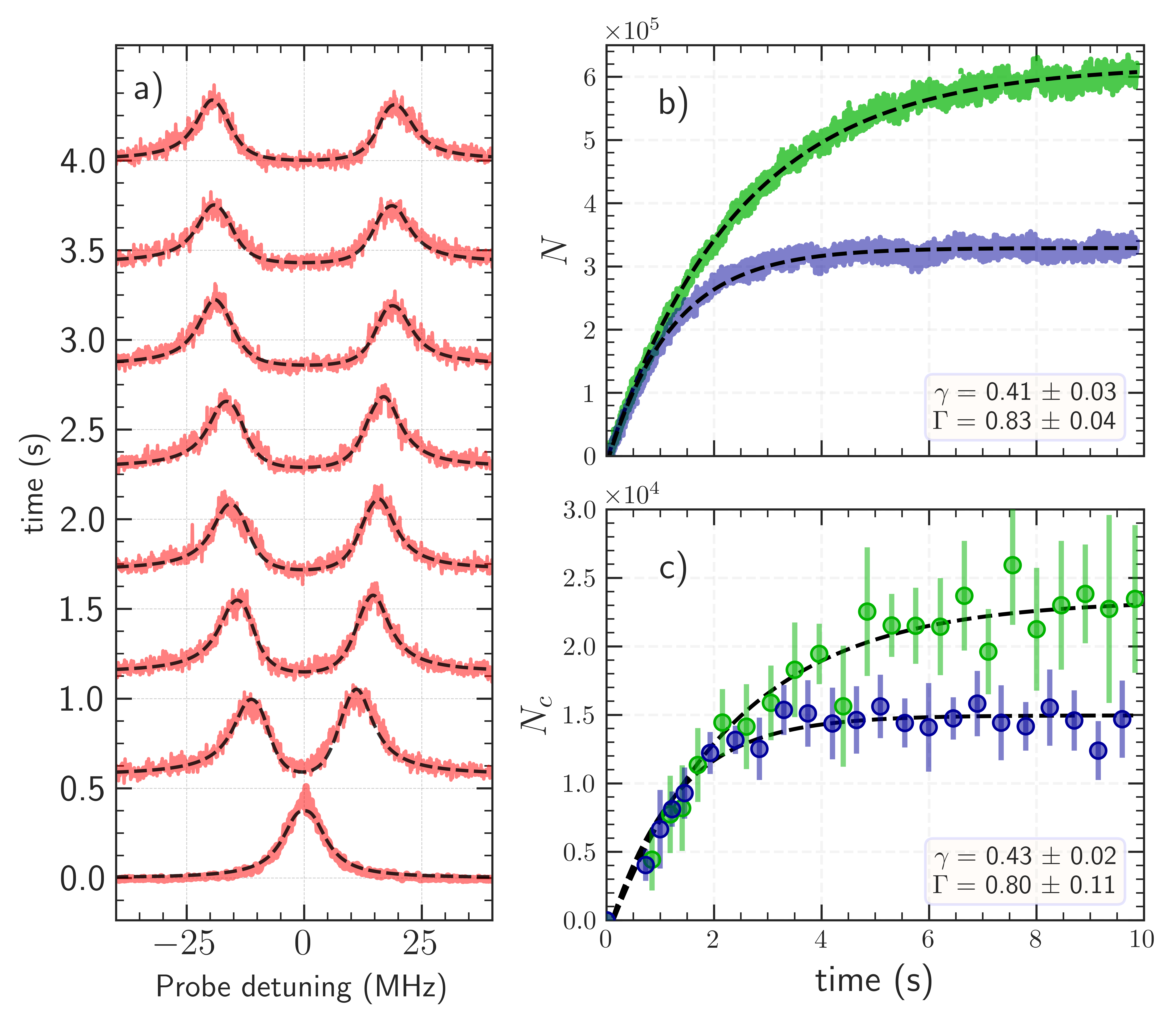} %
    \caption{Loading curve for bright MOT a) time evolution of VRS as the MOT loads b) loading curve from MOT fluorescence. Here, the green curve is when PA is off-resonant, and the blue curve is when PA is on-resonance. The loading rate $\Gamma$ and $\gamma$ are calculated from the fit. c) loading curve measured from VRS. Each point is an average of 5 repeated measurements and the error bar is the standard deviation. All the measurements are done on pa5 transition.}%
    \label{fig:bmot}%
\end{figure}

\begin{figure*}[ht]
    \centering
    \includegraphics[width=0.97\textwidth,height=8cm]{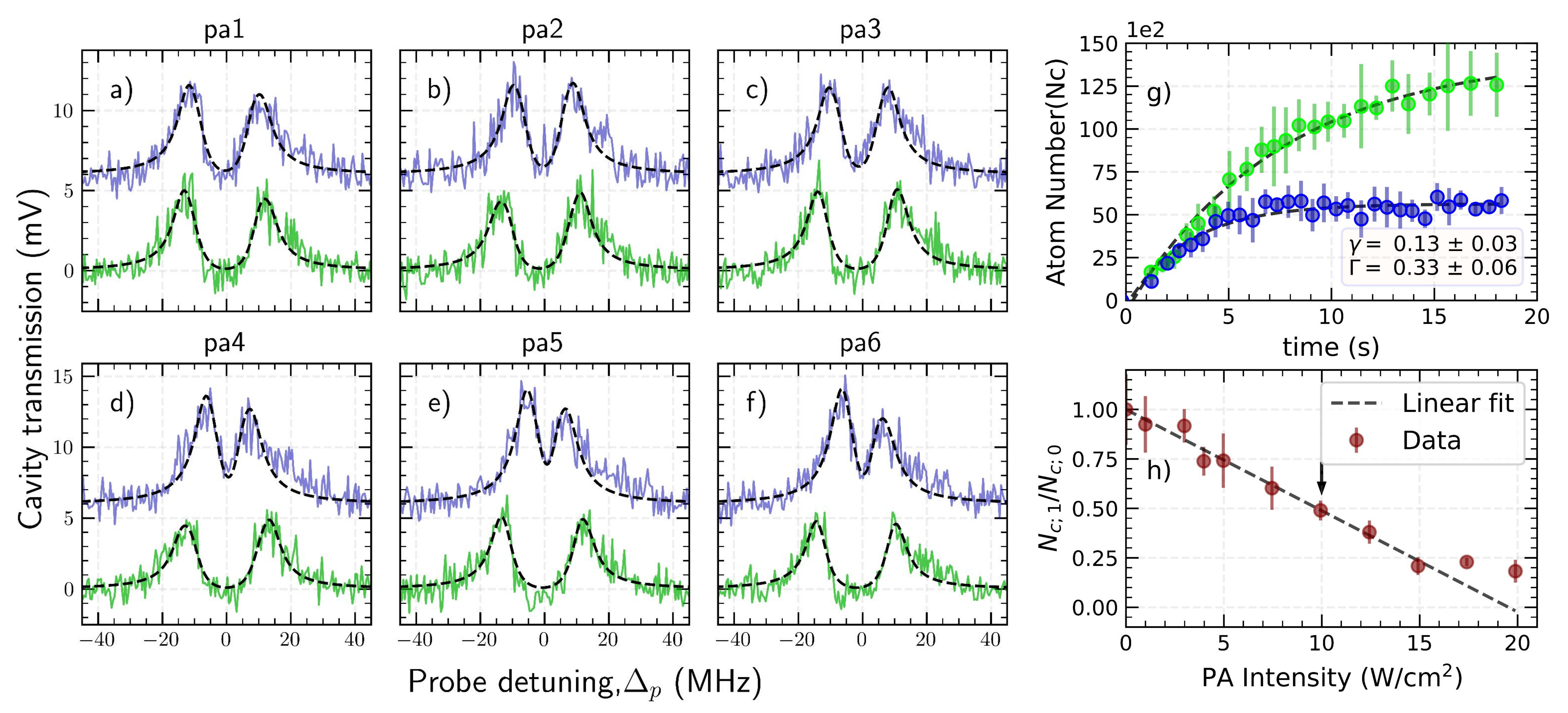}
    \caption{Direct detection of PA resonances in dark MOT using VRS. Here,(a)-(f) shows the change in VRS for various PA resonances. The blue line is when the PA laser is kept on-resonance, and the green plot is when the PA laser is kept off-resonant. The figure shows the VRS for a single sweep and the theoretical fit of the data to determine the peak separation. (g) Loading curve using atoms coupled to cavity mode(N${}_c$). Here, the green circle is when the PA laser is kept off-resonant to the transition, and the blue circle is when the PA laser is kept on-resonant. (h) The ratio of $N_c$ when PA laser is on-transition($N_{c;1}$) to off-transition($N_{c;0}$) as a function of PA intensity. The arrow represents the peak intensity used for dark MOT loading measurements. Each point is an average of 5 measurements. All the measurements are done on the pa5 transition.}
    \label{fig:dmot}
\end{figure*}

The MOT fluorescence with a bright MOT provides a direct calibration of the VRS detection. As bright MOT atoms are predominantly in $F=3$ state, the cavity is locked to $F=3$ to $F^{\prime}=3$ transition, and the probe is scanned across the same transition. For this measurement, value of $\tilde{g_0} = \sqrt{\mu_{33}^2\omega_{33}/{2\hbar\epsilon_{0}V_c}} \approx 0.13$ MHz. The frequency of PA transition is monitored by a wavemeter. Change in the VRS signal is observed in addition to the trap-loss fluorescence signal on PA resonances. The ground state molecules and majority of the free atoms formed from excited state molecules during a PA are not trapped in a MOT~\cite{STWALLEY1999194}, resulting in a reduction of atom number in MOT. This results in a smaller VRS when the PA laser is resonant to any of the transitions. This reduction in VRS measures the formation of excited state Rb${}_2$ molecules by PA.

As shown in fig~\ref{fig:bmot}, a direct measure of the loading curve using VRS is also performed for bright MOT with PA laser kept off and on transition. Here, the probe laser is scanned at a rate of 8 Hz for 10 seconds and frequency is referenced with respect to saturated absorption spectroscopy (SAS). The MOT coil is turned off for a short time, and the probe output is continuously monitored. As a result, once the MOT starts to load rapid VRS measurements are possible. The loading rate is calculated from the VRS signal. Fig~\ref{fig:bmot} a) shows the evolution of VRS as the MOT loads. Initially, when there are no atoms in the MOT, the cavity transmission corresponds to empty cavity transmission. As MOT builds up, the number of atoms in the cavity mode increases, resulting in an increase in VRS. 

Fig.~\ref{fig:bmot} (b) shows the MOT loading from fluorescence when the PA laser is kept on-resonance (blue) and off-resonance (green). This is fitted with $N(t) = N_0 \left(1-\exp\left({-\Gamma t}\right)\right)$ (see appendix) to extract loading rate ($\Gamma$). When the PA laser is kept off-resonance, the loading rate is $\gamma = 0.41 \pm 0.03$ s${}^{-1}$. Similarly, loading rate with on-resonant PA is $\Gamma = \gamma+\gamma_{PA} = 0.83 \pm 0.04$ s${}^{-1}$. The loss rate in MOT due to PA is therefore as $\gamma_{PA} = 0.42 \pm 0.07$ s${}^{-1}$. Fig.~\ref{fig:bmot} (c) shows the same loading measured using VRS with on-resonant (blue circle) and off-resonant (green circle) PA. From the fit, the loading rate with off-resonant PA is $\gamma  = 0.43 \pm 0.02$ s${}^{-1}$ and with on-resonant PA is $\Gamma = 0.80 \pm 0.11$ s${}^{-1}$. Hence, the loss rate in MOT calculated from VRS is $\gamma_{PA} = 0.37 \pm 0.13$ s${}^{-1}$. These values agree very well with MOT fluorescence measurement, validating VRS-based measurements.

\subsection{\label{pa_dmot} PA resonances and loading rates in dark MOT}

In systems like dark MOT, most of the atoms are in non-fluorescing state ($F = 2$). As a result, a direct fluorescence trap-loss signal for PA resonances is not possible. The VRS measurement makes such detection possible even when the system is in a dark state. For this detection, the cavity is locked to $F=2$ to $F'=3$ transition. A weak probe laser is scanned across the same transition and referenced to SAS. Since the atoms are in $F =2 $ state, all the PA resonances (pa1-pa6) are shifted by $\sim 3.03$ GHz, which is the difference in frequency between $F=2$ and $F=3$ hyperfine levels. The PA laser with peak intensity $\sim 15$ W$/$cm${}^2$ is incident on the dark MOT. The cavity transmission is monitored with PA kept on and off-resonant for all six transitions.

Figure~\ref{fig:dmot} (a)-(f) shows the change in VRS signal when the PA laser is on-resonance (blue) and off-resonance (green) for all the transitions. In all cases, the blue curve is shifted for better visibility. The values of all VRS and corresponding molecular transitions are given in the appendix (Table 1). As seen from fig~\ref{fig:dmot} the value of VRS decreases when a PA transition is addressed, which is a direct measurement of the formation of Rb${}_2^*$ molecule due to PA in a dark MOT.

To demonstrate the time evolution capabilities, the loading rates for dark MOT are measured in the presence of the PA. For this measurement, the probe laser is locked to $F=2$ to $F'=3$ transition using SAS. This locked laser is passed through two double-pass AOMs. One of the AOMs is scanned continuously at a rate of $4$ Hz for $20$ seconds. For this loading measurements peak intensity of the PA laser was kept $\sim 10$ W$/$cm${}^2$. Figure~\ref{fig:dmot} (g) shows the $N_c$ as a function of time when the PA laser is kept off-resonance (green circle) and on-resonance (blue circle). Loading rate of dark MOT on resonance, $\Gamma_D = \gamma_D + \gamma_{PA;D} = 0.33 \pm 0.06$ s${}^{-1}$, where $\gamma_D = 0.13 \pm 0.03 $ s${}^{-1}$, for pa5, as shown in fig~\ref{fig:dmot}(g). From this, the loss rate in dark MOT due to PA is calculated as $\gamma_{PA;D} = 0.20 \pm 0.09$ s${}^{-1}$. Figure~\ref{fig:dmot} (h) shows the ratio of $N_c$ when PA is on and off-resonant to the transition as a function of PA laser intensity. As the PA intensity increases the fraction of atoms getting converted to PA molecule increases, showing a reduction in the ratio of $N_c$ value. The dashed line, which is a linear fit shows good agreement for lower PA intensity to theoretical expectation. For higher intensity saturation effect results in deviation from the linear fit. This saturation in PA rate for higher PA intensity is well known both theoretically~\cite{Bohn99} and experimentally~\cite{Drag,Prodan,Zimmermann}.

\subsection{\label{pa_temp}Temperature measurement during PA}
\begin{figure}[t!]
    \centering
    \includegraphics[width=0.48\textwidth]{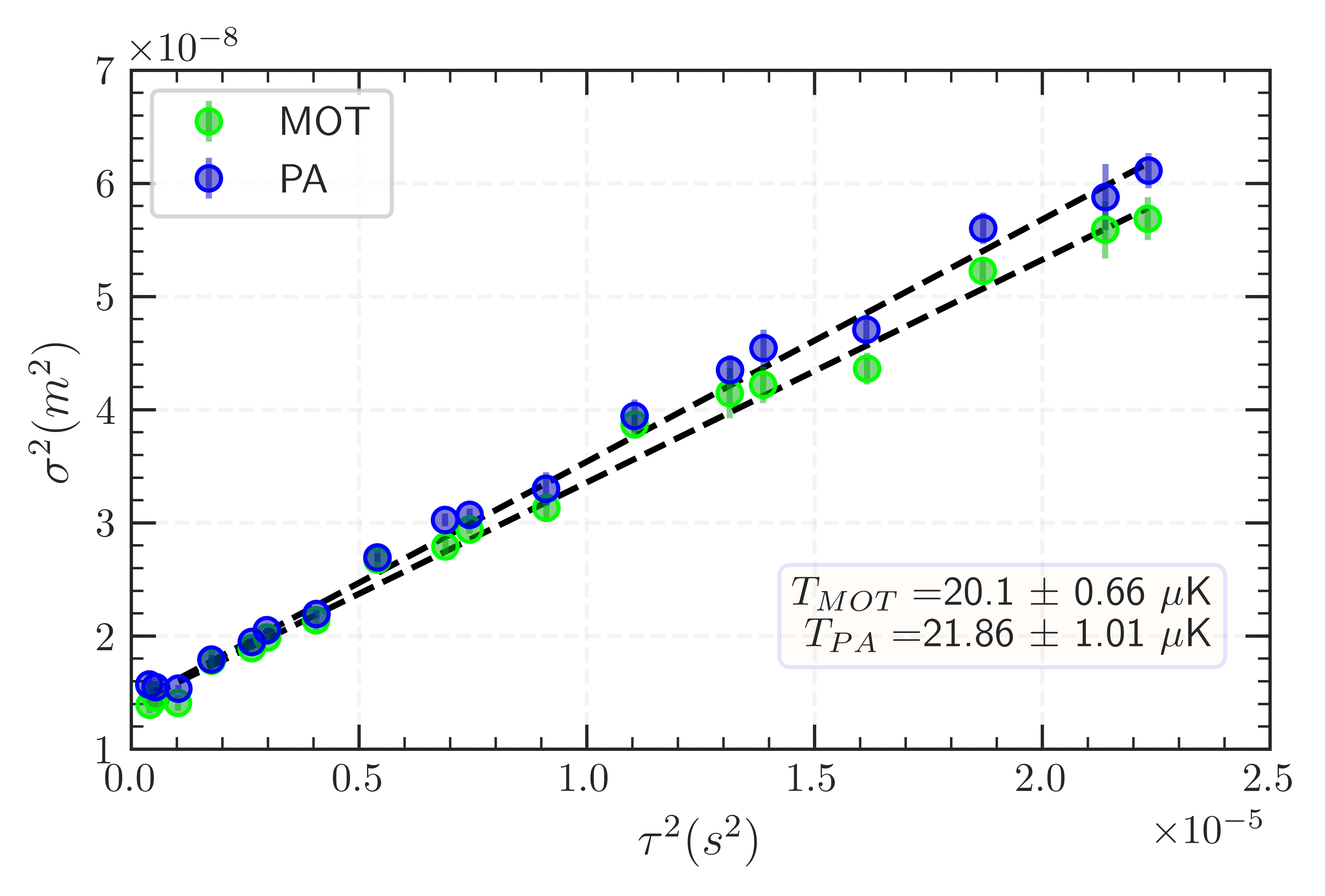}
    \caption{Temperature measurement during PA. Here, the green circle represents measurement with PA, and the blue circle without PA. Each data point is an average of 5 measurements.}
    \label{fig:fig6}
\end{figure}
A small fraction of atoms that are dissociated from excited molecules can have kinetic energy less than the trap depth. As a result, these atoms don't contribute to the trap-loss signal. However, if they have sufficient energy to increase the average temperature of the MOT~\cite{Walhout}, an extension of this detection technique is possible. To demonstrate this, temperature measurement using cavity VRS~\cite{Tridib} is done for bright MOT with and without PA. The cavity is locked to $F=3 \rightarrow F'=4$ transition. The probe laser is locked to the same transition, and AOM is scanned at a rate of $2$ kHz. MOT beams are switched off using an AOM-based switch for $5$ ms. During this period, MOT expands ballistically, and from the measured value of $N_c$ the standard deviation $(\sigma)$ of MOT can be calculated. This gives a direct measure of temperature using the expression~\cite{Weiss}, $\sigma(\tau)^2=\sigma_0^2 + \frac{k_B T}{m} \tau^2$. The same procedure is repeated with an on-resonant PA, which is turned off just before the MOT expansion. This ensures that the atoms that are recaptured in the trap contribute to the temperature of the MOT.  As shown in fig~\ref{fig:fig6}, the measured value of temperature of the MOT is $T_{MOT} = 20.10 \pm 0.66$ $\mu$K and with on-resonant PA is $T_{PA} = 21.86 \pm 1.01$ $\mu$K, which shows that fraction of atoms recaptured by the decay of the excited molecules is negligible.

\section{\label{sec:con} Discussion \& Conclusion}
The advantages of PA from an optically dark ensemble are numerous, the most prominent is the larger fraction of atoms converted to molecules (fig.\ref{fig:dmot}), since the dark MOT atoms are all in the same ground state. Further in systems like dark MOT, the PA rate $\left(\gamma_{PA;D}\right)$ is larger compared to the loading rate $\left( \gamma_D \right)$, giving a higher fraction of PA (see appendix). The cavity-based technique of detection which probes the free atoms instead of the molecules directly~\cite{sawant,zhu2020}, can be used very effectively for detection, in scenarios where detection is difficult and in cases like 3-particle PA~\cite{Schnabel} or PA of Rydberg molecules~\cite{Anderson} with accuracy. 

This technique is expected to be particularly useful where direct fluorescence detection is not possible. As the method is non-destructive by nature, atom-molecule population dynamics in the system can be studied continuously. Since ultracold molecules are not trapped in our experiment, a direct detection of atom-molecule collision was not possible. However, the extension of this method for such experiments is straightforward, making this way of studying interactions more universal. In summary, the present work demonstrates how cavity techniques can be applied to detect complex processes and yield accurate results in very challenging systems and spatially compact geometries.

\begin{acknowledgments}
The authors would like to thank the Department of Science and Technology and Ministry of Electronics and Information Technology (MeitY), Government of India, under a Centre for Excellence in Quantum Technologies grant with Ref. No. 4(7)/2020-ITEA.
\end{acknowledgments}

\appendix

\section{Loading rates in MOT}

The rate equation for the number of atoms ($N$) in the trap with PA laser is given by~\cite{Hoffmann94,Drag} 
\begin{equation}\label{eq1}
    \frac{dN}{dt} = L - \gamma N -\left(\beta+\beta_{PA}\right) \int n^2(r) d^{3} r
\end{equation}
where, L is the rate at which atoms are loaded into the trap, $\gamma N$ is the rate at which atoms are lost from the trap due to background collisions, $\beta$ is the loss rate due to collision with trapped atoms, and $\beta_{PA}$ is the loss rate due to photoassociation. Dark MOT has a uniform density~\cite{Niranjan} $\left(n_{max}\right)$ inside the trap. As a result,  eq~\ref{eq1} can be rewritten as 
\begin{equation}\label{eq2}
    \begin{split}
        \frac{dN}{dt} &= L - \left(\gamma +\left(\beta+\beta_{PA}\right) n_{max}\right) N \\
        N(t) &= N_0 \left(1-\exp\left({-\Gamma t}\right)\right)
    \end{split}
\end{equation}
where, $\Gamma = \gamma + \left(\beta+\beta_{PA}\right) n_{max}$. In steady state, 
\begin{equation}\label{eq3}
    \frac{N_{PA}}{N_{at}} = 1 - \frac{\gamma+\beta n_{max}}{\gamma + \left(\beta+\beta_{PA}\right) n_{max}}    
\end{equation}
where, $N_{at}$ is the steady state atom number without PA and $N_{PA}$ is the fraction of atoms converted into PA. Even for a bright MOT with peak density, $\rho_0 \sim 8 \times 10^{10}$ cm${}^{-3}$ this can be used to find the approximate value of $\beta_{PA}$ assuming a radiation trap model for MOT density~\cite{Hoffmann94}.
\medskip
\section{PA resonances in dark MOT}
\begin{table}[t!]
    \centering
    \begin{tabular}{m{1cm}  m{2.5cm}  m{1.5cm} m{1.5cm} }
        \hline\\[0.1cm]
        & \multicolumn{1}{p{2.5cm}}{\centering Nearest PA \\ Resonance}
        & \multicolumn{1}{p{1.5cm}}{\centering VRS(off) \\ MHz} 
        & \multicolumn{1}{p{1.5cm}}{\centering VRS(on) \\MHz} \\[0.5cm]\hline \\
        \multicolumn{1}{p{1cm}}{\centering pa1 }
        & \multicolumn{1}{p{2.5cm}}{\centering $0_g^{-} , v=50$ \\ $1_g , v=164$ \\ $0_u^{+} ,v=251$}
        & \multicolumn{1}{p{1.5cm}}{\centering$25.41$ \\ $\pm 0.8$}
        & \multicolumn{1}{p{1.5cm}}{\centering$22.53$ \\$\pm 1.01$ } \\[1cm]

        \multicolumn{1}{p{1cm}}{\centering pa2 }
        & \multicolumn{1}{p{2.5cm}}{\centering $0_g^{-} , v=51$ \\ $1_g , v=165$ \\ $0_u^{+} ,v=252$}
        & \multicolumn{1}{p{1.5cm}}{\centering$24.43$ \\ $\pm 0.54$ } 
        & \multicolumn{1}{p{1.5cm}}{\centering$20.95$ \\ $\pm 1.44$ } \\[1cm]

        \multicolumn{1}{p{1cm}}{\centering pa3} 
        & \multicolumn{1}{p{2.5cm}}{\centering $0_g^{-} , v=52$ \\ $1_g , v=166$ \\ $0_u^{+} ,v=253$}
        &\multicolumn{1}{p{1.5cm}}{\centering $24.82$ \\ $\pm 1.141$ } 
        &\multicolumn{1}{p{1.5cm}}{\centering $19.33$ \\ $\pm 1.9$ } \\[1cm]

        \multicolumn{1}{p{1cm}}{\centering pa4 } 
        & \multicolumn{1}{p{2.5cm}}{\centering $0_g^{-} , v=53$ \\ $1_g , v=168$ \\ $0_u^{+} ,v=255$}
        &\multicolumn{1}{p{1.5cm}}{\centering $22.14$ \\ $\pm 1.19$ }
        &\multicolumn{1}{p{1.5cm}}{\centering $15.18$ \\ $\pm 1.66$ } \\[1cm]

        \multicolumn{1}{p{1cm}}{\centering pa5  }
        & \multicolumn{1}{p{2.5cm}}{\centering $0_g^{-} , v=54$ \\ $1_g , v=169$ \\ $0_u^{+} ,v=256$}
        &\multicolumn{1}{p{1.5cm}}{\centering $25.38$ \\ $\pm 1.3$ }
        &\multicolumn{1}{p{1.5cm}}{\centering $12.13$ \\ $\pm 0.58$ } \\[1cm]

        \multicolumn{1}{p{1cm}}{\centering pa6} 
        & \multicolumn{1}{p{2.5cm}}{\centering $0_g^{-} , v=58$ \\ $1_g , v=173$ \\ $0_u^{+} ,v=260$}
        &\multicolumn{1}{p{1.5cm}}{\centering $24.28$ \\ $\pm 1.32$} 
        &\multicolumn{1}{p{1.5cm}}{\centering $12.7$ \\$\pm 0.46$ } \\[1cm]

        \hline
    \end{tabular}
    \caption{VRS values for different PA resonances. VRS(off) is when the PA laser is off-resonant, and VRS(on) is when the PA laser is on-resonant.}
    \label{tab:1}
\end{table}
The table~\ref{tab:1} shows the values of VRS measured for different PA transitions in dark MOT. Here, VRS (off) represents the vacuum Rabi split when the PA laser is kept off-resonance, and VRS (on) is when the PA laser is kept on-resonance. The data is fitted with a theoretical expression for intra-cavity intensity derived from a two-level atom coupled to a cavity to extract VRS values. Each value given in the table is an average of 10 measurements, and the error bar is the standard deviation. As seen from the table~\ref{tab:1} the average value of VRS is always smaller when PA resonance is addressed. The fraction of excited Rb${}_2$ molecules created can be calculated from $1-\left(N_{c;1}/N_{c;0}\right) = 1- \left(VRS_1{}^2/VRS_0{}^2\right)$, where $VRS_1$ is the split when PA is on resonance and $VRS_0$ is corresponding off-resonance value. All the PA transitions are identified from ab initio calculation done in ref~\cite{jyothi}.
 
\newpage
\bibliography{ref.bib}

\end{document}